\documentclass[sigconf]{acmart} 
\AtBeginDocument{%
  }

\usepackage{stfloats}   
\usepackage{makecell} 
\usepackage{soul}
\usepackage{balance}

\copyrightyear{2026}
\acmYear{2026}
\setcopyright{cc}
\setcctype{by-nc-nd}
\acmConference[CHI '26]{Proceedings of the 2026 CHI Conference on Human Factors in Computing Systems}{April 13--17, 2026}{Barcelona, Spain}
\acmBooktitle{Proceedings of the 2026 CHI Conference on Human Factors in Computing Systems (CHI '26), April 13--17, 2026, Barcelona, Spain}
\acmPrice{}
\acmDOI{10.1145/3772318.3791339}
\acmISBN{979-8-4007-2278-3/2026/04}

\begin{document}
\settopmatter{printfolios=true} 


\newcommand{\systemname}{HiFiGaze}
\newcommand{\vm}[1]{\textcolor{blue}{[Vimal: #1]}}
\newcommand{\tj}[1]{\textcolor{red}{[Taejun: #1]}}

\definecolor{pink}{RGB}{0,0,0}
\newcommand{\new}[1]{\textcolor{pink}{#1}}

\title{\systemname: Improving Eye Tracking Accuracy Using Screen Content Knowledge}

\author{Taejun Kim}
\affiliation{%
  \department{Human-Computer Interaction Institute}
  \institution{Carnegie Mellon University}
  \city{Pittsburgh}
  \state{Pennsylvania}
  \country{USA}}
\email{taejunkim.me@gmail.com}

\author{Vimal Mollyn}
\affiliation{%
  \department{Human-Computer Interaction Institute}
  \institution{Carnegie Mellon University}
  \city{Pittsburgh}
  \state{Pennsylvania}
  \country{USA}}
\email{vmollyn@cs.cmu.edu}

\author{Riku Arakawa}
\affiliation{%
  \department{Human-Computer Interaction Institute}
  \institution{Carnegie Mellon University}
  \city{Pittsburgh}
  \state{Pennsylvania}
  \country{USA}}
\email{rarakawa@andrew.cmu.edu}

\author{Chris Harrison}
\affiliation{
  \department{Human-Computer Interaction Institute}
  \institution{Carnegie Mellon University}
  \city{Pittsburgh}
  \state{Pennsylvania}
  \country{USA}}
\email{chris.harrison@cs.cmu.edu}

\renewcommand{\shortauthors}{Kim et al.}

\begin{abstract}
We present a new and accurate approach for gaze estimation on consumer computing devices. We take advantage of continued strides in the quality of user-facing cameras found in e.g., smartphones, laptops, and desktops --- 4K or greater in high-end devices --- such that it is now possible to capture the 2D reflection of a device's screen in the user's eyes. This alone is insufficient for accurate gaze tracking due to the near-infinite variety of screen content. Crucially, however, the device knows what is being displayed on its own screen --- in this work, we show this information allows for robust segmentation of the reflection, the location and size of which encodes the user's screen-relative gaze target. We explore several strategies to leverage this useful signal, quantifying performance in a user study. Our best performing model reduces mean tracking error by \textasciitilde18\% compared to a baseline appearance-based model. A supplemental study reveals an additional 10-20\% improvement if the gaze-tracking camera is located at the bottom of the device. 
\end{abstract}



\begin{teaserfigure}
  \centering
  \includegraphics[width=1.0\linewidth]{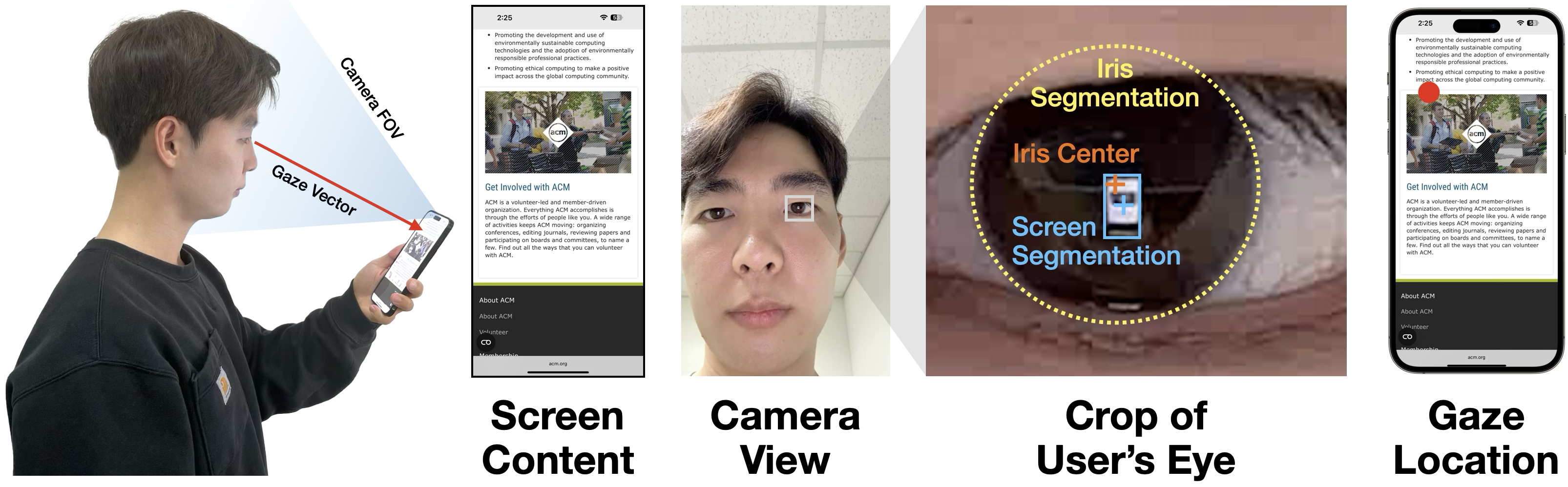}
  \vspace{-.6cm}
  \caption{HiFiGaze leverages two pieces of information: 1) crop of the user's eyes (as captured by increasingly common high-quality, 4K "selfie" cameras), and 2) the contents of the screen (which the device already knows). We quantify how combining these two information streams improves eye tracking accuracy over a conventional appearance-based approach.}
  \vspace{.4cm}
  \label{fig:hero}
  \Description{This figure shows the overall HiFiGaze pipeline. A user looks at a smartphone, where both the front-facing camera image of the face and the screen content are captured. The eye region is cropped and analyzed together with screen information to estimate the gaze point on the display.}
\end{teaserfigure}

\maketitle

\section{Introduction}

\begin{figure*}[b]
	\centering \includegraphics[width=1.0\linewidth]{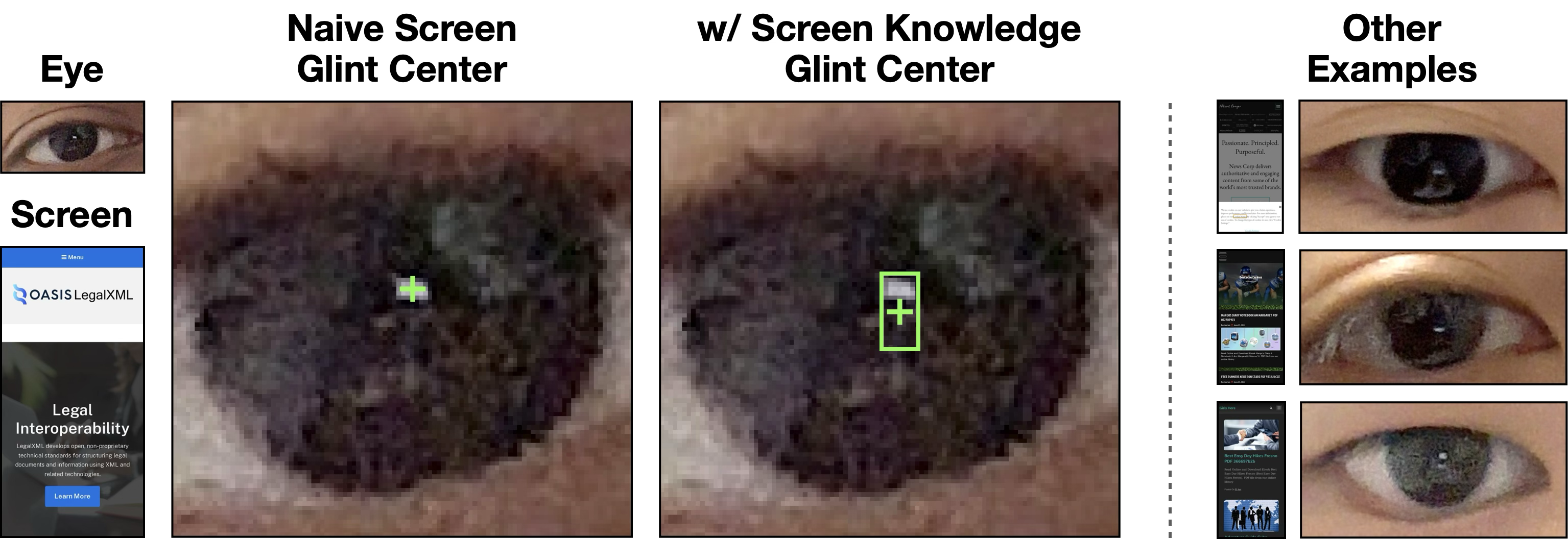}
	\caption{An example smartphone screen drawn from the WebUI dataset~\cite{wu2023webui}. A naive screen glint approach will detect the light element, but incorrectly estimate the reflection center. With knowledge of the screen content, HiFiGaze finds the correct screen reflection center, necessary for accurate gaze estimation. Right: additional examples where a naive screen glint center approach would fail. }
	\label{fig:NaiveGazeErrors}
    \Description{This figure illustrates why naïvely using the brightest reflection in the eye can fail. Bright elements in the screen content can shift the detected glint away from the true screen center. Using screen content knowledge enables more accurate identification of the correct reflection corresponding to the display.}
\end{figure*}

High-accuracy gaze tracking generally requires specialized devices, such as glasses/headsets with integrated close-range cameras~\cite{huang2024measuring,wei2023preliminary,ehinger2019new} (offering a stable vantage point and high-resolution capture of the eye) and/or infrared illuminators enabling geometric model-based estimation~\cite{guestrin2006general,housholder2021evaluating}. Of course, it would be most desirable if gaze tracking could be enabled on commonplace computing devices, such as smartphones, laptops, and desktops. Indeed, prior work~\cite{krafka2016eye,zhang2017s,valliappan2020accelerating} has demonstrated this is possible using the built-in, user-facing cameras available in these devices, but these appearance-based methods tend to provide only modest accuracy (e.g., mean offsets of \textasciitilde2 cm without per-user calibration~\cite{valliappan2020accelerating,DBLP:conf/icmi/ArakawaG0A22,huynh2021imon}).

In this work, we introduce a new gaze estimation method --- HiFiGaze --- that operates akin to a model-based approach, but runs on commonplace computing devices using their existing cameras. The core insight of our work is that many devices now offer 4K (or greater) user-facing camera streams. This means we can not only capture the appearance of the eye, but also the reflection of the device's screen in the eye~(Figure~\ref{fig:hero}). However (and as we will show in our evaluation), this data is insufficient to accurately resolve the gaze vector. The main culprit is the near-infinite variety of content that can be shown on a device's screen, which can lead to misalignment if looking for a screen "glint" (see example in Figure~\ref{fig:NaiveGazeErrors}). Fortunately, the device knows what is being displayed on its own screen, and this extra knowledge allows our method to robustly segment screen reflections irrespective of the content (there are some specific cases where it fails, which we discuss). Once segmented, the location and size of the screen reflection relative to the pupil center directly encodes the gaze vector. 

We note that we are not the first to utilize screen reflections of a computer display on the eye. Most notably, ScreenGlint~\cite{huang2017screenglint} also employed a smartphone's user-facing camera to estimate the point of gaze on the screen. However, this work did not utilize screen content knowledge, and its low-resolution camera meant the screen reflection only appeared as a small 2D glint point. Most importantly, the method only functioned when the screen was essentially a full white field (to create the glint), which is not common or practical. 


There are several ways to fuse high-resolution eye images and screen content, and we evaluate seven promising encodings alongside a baseline appearance-based method. The most successful method uses eye crops and reflection vectors (Figure~\ref{fig:architectures}), and achieves a mean gaze accuracy of 1.64 cm on an iPhone 14 Pro Max (without any per-user calibration). 
We also found long eyelashes and upper eyelids can sometimes occlude screen reflections, particularly when the user fixates on targets near the bottom of the screen. We hypothesized that a camera operating at the bottom of the smartphone might sidestep this issue, and ran a supplemental study to quantify its effect on performance. 
Taken together, our investigations and results underscore that app-icon-sized, calibration-free gaze accuracy is within reach for commonplace computing devices, without the need for additional hardware (sensors, illuminators, etc.) like those found in specialized gaze systems.

\section{Related Work}
Gaze tracking methods and interaction techniques built on top of them have been extensively studied. We refer readers to comprehensive surveys by Cheng et al.~\cite{cheng2024appearance} and Lei et al.~\cite{lei2023end} for a summary of the field and fundamental techniques. In this section, we provide a primer on the main approaches, chiefly to underscore the unique contributions of this work. 

Unlike prior work~\cite{DBLP:conf/icmi/ArakawaG0A22,kong2021eyemu,kim2019nvgaze,du2022freegaze,abdelrahman2023l2cs}, which divide their literature review by appearance-based and model-based approaches, we instead split our discussion along a more practical dimension --- whether or not additional hardware is needed, most notably dedicated cameras and/or illuminators. We start with systems that require specialized hardware not found in laptops, smartphones, and similar computing devices. We then move to methods more similar to our own, that can track the eyes using standard RGB cameras, already built into many computing devices. We conclude this section with a more ranging review of other computing systems that have leveraged high-resolution imagery of the eyes.

\subsection{Gaze Estimation Using Specialized Trackers}

Wearable gaze tracking systems such as glasses~\cite{tobiiProducts,pupilLabsNeon} and XR headsets~\cite{appleVisionPro,metaQuestPro,viveElite}  typically embed eye-facing infrared (IR) illuminators and cameras. The IR illumination of the eye produces reflections on the cornea (glints) which can be used to track gaze. The most common technique is Pupil Center Corneal Reflection (PCCR), defined as a vector from the pupil's center to the glint. 
Since the pupil shifts with eye rotation while the glints remain relatively stationary, PCCR provides strong cues for gaze estimation. This signal is then processed by either geometric models or hybrid methods with neural networks~\cite{hansen2009eye}. The Apple Vision Pro, for instance, shows robust gaze tracking performance with a mean accuracy of less than 1\textdegree{}~\cite{huang2024measuring}. There are also many commercially available eye tracking peripherals; the Tobii Eye Tracker 5~\cite{tobiiProducts}, for example, has an accuracy of about \textasciitilde0.8\textdegree{} in unconstrained settings~\cite{housholder2021evaluating}. 
While these systems have been tested and verified in real-world consumer settings, they rely on dedicated hardware to provide high-accuracy gaze estimation.

\subsection{Gaze Estimation on Commodity Devices}

Gaze tracking on commodity devices, such as smartphones, tablets, and laptops, has been actively studied for the past decade. Seminal work by Krafka et al.~\cite{krafka2016eye} introduced a large-scale dataset containing 2.5 million frames with ground truth gaze annotations, collected from 1,450 mobile device users in unconstrained settings. Using a deep neural network trained on this dataset, they demonstrated gaze accuracy of 1.9~cm. Since then, a substantial body of work~\cite{zhang2015appearance,huang2017tabletgaze,huang2017screenglint,zhang2017s,guo2019generalized,he2019device,bao2021adaptive,valliappan2020accelerating,huynh2021imon,DBLP:conf/icmi/ArakawaG0A22,gunawardena2025smartphone} has explored new ways to further improve performance. The most common input is an eye-region patch cropped from an RGB camera image via facial landmark detection~\cite{huang2017tabletgaze, guo2019generalized, bao2021adaptive, lei2023dynamicread, namnakani2023comparing, lei2025quantifying, zhong2024uncertainty}.
Researchers have also explored including auxiliary features, such as head pose~\cite{zhang2015appearance}, face grid maps~\cite{krafka2016eye}, and eye corner landmarks~\cite{he2019device,valliappan2020accelerating,huynh2021imon}, in order to provide additional cues of the relative displacement between the user and the camera. Concurrently, considerable research has been invested in designing more effective architectures and evaluating them on popular datasets such as GazeCapture~\cite{krafka2016eye}, TabletGaze~\cite{huang2017tabletgaze}, and MPIIGaze~\cite{zhang2015appearance}. 

However, it is important to note that these datasets were collected roughly a decade ago, with user-facing cameras with resolutions limited to around 1280$\times$720 (0.9 MP). In this work, we highlight a new opportunity enabled by the continued advancement of user-facing cameras in consumer devices, many of which now provide 4K resolution (8.3 MP) or higher. At such resolutions, it is possible to capture the reflection of a device's screen in the user's eyes. While the variety of screen content makes it difficult for conventional models to exploit such reflections, our approach leverages the knowledge of what is being displayed on the screen. This allows our method to robustly segment the screen reflection within the eye, independent of content, and integrate this useful signal into prediction.

\subsection{Uses of High-Resolution Eye Capture}

Zooming out from gaze estimation, there are other computing systems that leverage high-resolution eye imagery for various purposes, as summarized by Nitschke~et al.~\cite{DBLP:journals/imt/NitschkeNT13}.
In 2004, Nishino and Nayar~\cite{nishino2004world} discussed the wealth of visual information that a single eye image can contain. Using a 6 MP Kodak 760 digital camera, they considered applications including environment reconstruction, object recognition, human affect analysis, gaze estimation, and relighting~\cite{DBLP:journals/tog/NishinoN04}.
Jenkins and Kerr~\cite{Jenkins2013Identifiable} demonstrated that high-resolution portraits could reveal identifiable images of bystanders captured in corneal reflections, highlighting the forensic potential of eye imagery.
More recently, Alzayer et al.~\cite{alzayer2024seeing} introduced a system that could roughly reconstruct the 3D scene of the user's field of view from a series of portrait images of the user.

More in the HCI domain, ReflecTouch~\cite{zhang2022reflectouch} leveraged occluded patterns in screen reflections to infer the user's smartphone grasp posture. Schneider and Grubert \cite{schneider2017towards} explored how high-resolution eye images (captured by a 36 MP Sony Alpha 7r camera) could be used to support novel around-device interaction techniques (e.g., detecting objects and hands around the device) by analyzing the corneal reflection. Collectively, this line of work reflects the viability of treating the eye as a natural reflective surface rich in environmental information.

\section{Implementation} \label{implementation}
We now describe the implementation of \systemname, covering all components to provide a centralized overview. We A/B test many of these elements in our subsequent user studies to independently assess their efficacy.

\subsection{Test Hardware \& Study Apparatus} 
\label{sec:hardware}
While our approach is applicable to any computer device with a screen (see example reflections from different devices in Figure~\ref{fig:ExampleDevices}), in this work we selected a smartphone as a popular and representative device. For this, we used an iPhone 14 Pro Max. This device features a 12 MP user-facing ("selfie") camera with the capability to record and interactively process 4K (8.3 MP) video. We note that a few days before submission of this paper, Apple launched the iPhone 17 with an upgraded 18 MP user-facing camera.

\subsection{Camera Preprocessing} 

\begin{figure}[t]
	\centering \includegraphics[width=0.95\linewidth]{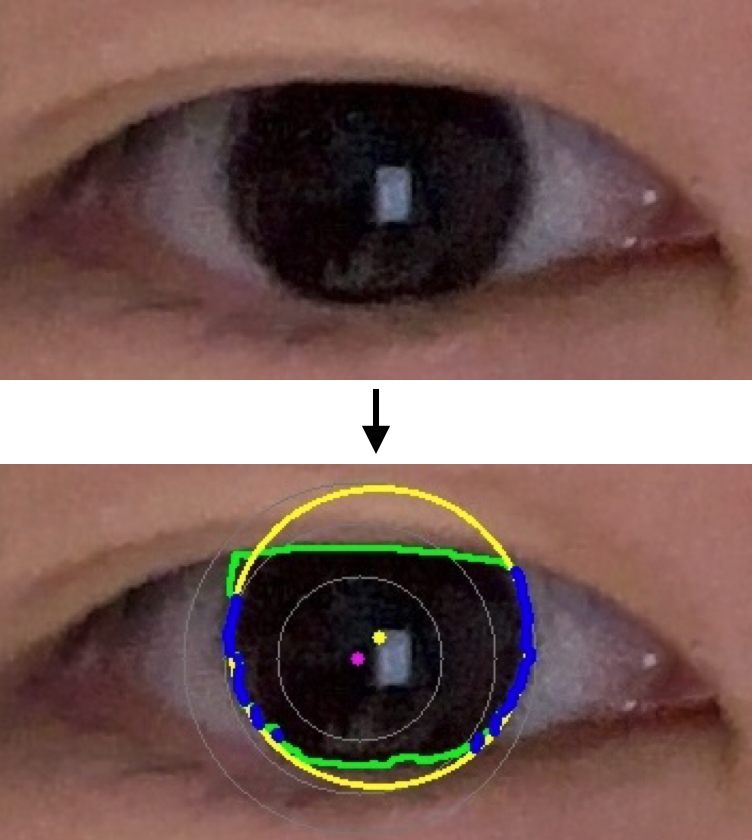}
	\caption{Iris center estimation. The purple dot shows the initial center estimate from MediaPipe~\cite{lugaresi2019mediapipe}, which can be noisy. To correct this estimate, we extract the iris contour (shown in green) using the GrabCut algorithm~\cite{rother2004grabcut}, then apply RANSAC~\cite{fischler1981random} to reject outliers before circle fitting. The blue dots show the final points used for fitting, and the yellow circle with its center dot shows the fitted result.}
	\label{fig:iris_center}
    \Description{This figure depicts the iris center refinement process. An initial iris center estimate is obtained, followed by iris segmentation and contour fitting. The final fitted circle produces a more accurate iris center used for gaze estimation.}
\end{figure}

The first step of our software pipeline is to locate the user's face and eyes in the camera's field of view~\cite{khamis2018understanding}. For this, we use MediaPipe’s face landmark detector~\cite{lugaresi2019mediapipe}. MediaPipe ~\cite{lugaresi2019mediapipe} outputs include the iris center landmarks. However, we found its predictions to be unacceptably noisy for our needs (Figure~\ref{fig:iris_center}). For this reason, we implemented a custom correction step on top of MediaPipe's outputs. Following the approach in ~\cite{huang2017screenglint}, we apply the GrabCut algorithm~\cite{rother2004grabcut} for iris segmentation and then perform circle fitting for a more stable detection. Specifically, we define the GrabCut regions as follows: r1 (definite iris foreground) is set to 0.4 times the iris height, r2 (probable foreground) to 0.5 times the iris width, and r3 (probable background) to 0.65 times the iris width, with 5 GrabCut iterations. We filter eyelid points from the segmented contours by excluding points where the local contour angle relative to the horizontal is less than 45\textdegree{}. For the remaining points, we apply RANSAC~\cite{fischler1981random} to remove outliers, and then finally, extract the iris center through geometric least-squares circle fitting~\cite{gander1994least}. The iris circle also gives us a robust estimate of iris diameter. We perform this process for both eyes.

\subsection{Model Input Encoding Strategies} 
\label{sec:EncodingStrategies}

\begin{figure*}[t]
	\centering
	\includegraphics[width=1.0\linewidth]{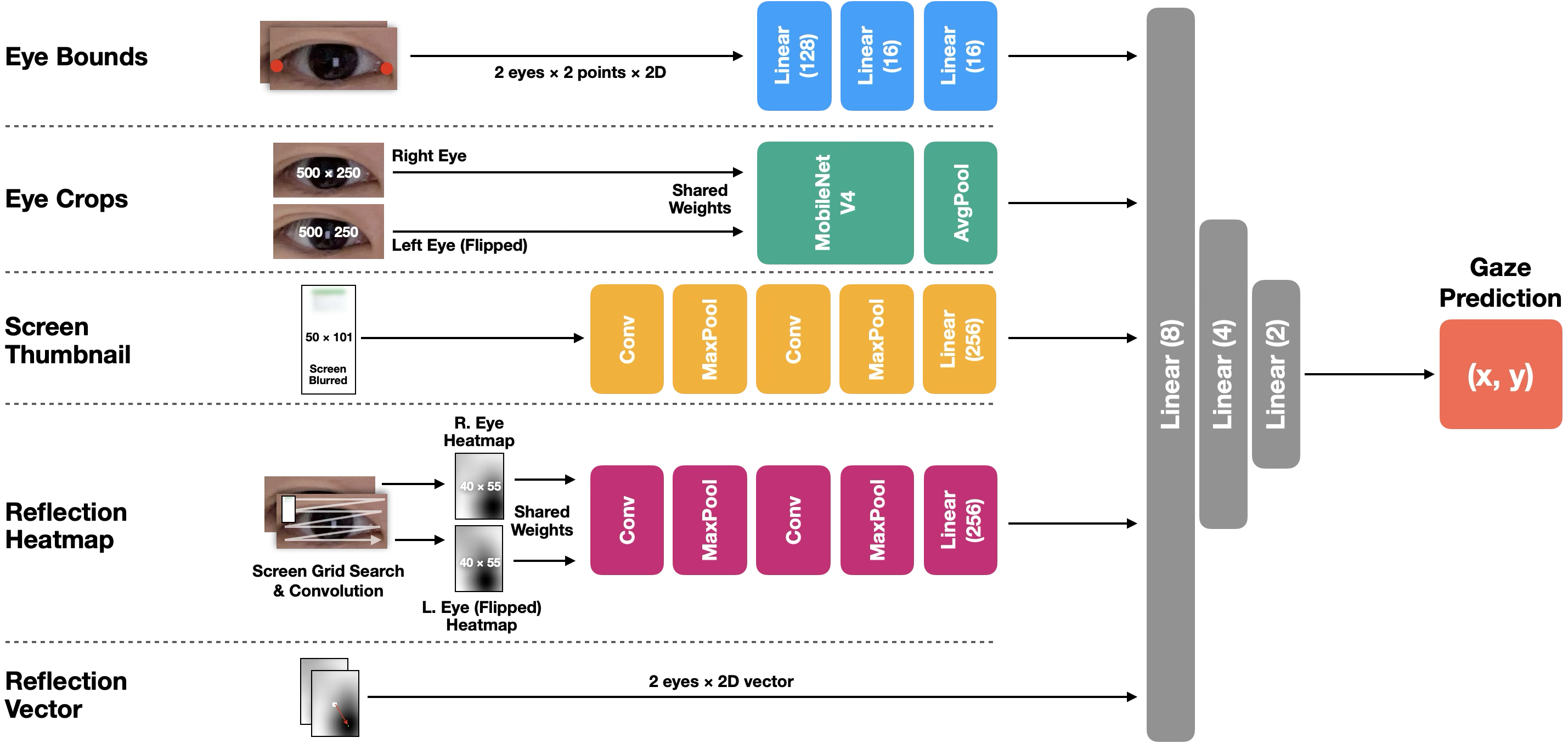}
	\caption{HiFiGaze architecture overview. Our model can draw on different sources of information, much of which leverages knowledge of the screen content. Our user study systematically compares the information power of these different inputs.}
	\label{fig:architectures}
    \Description{This figure presents the HiFiGaze model architecture. Multiple inputs, including eye crops, eye bounds, screen thumbnails, and reflection-based features, are encoded by separate modules. These features are fused to predict the final two-dimensional gaze location.}
\end{figure*}


We now describe five sources of information that we pass to our model, increasing in abstraction (illustrated in Figure~\ref{fig:architectures}). 

\subsubsection{Eye Bounds}
\label{sec:eyebounds}
The left and right corners of the eye (in the camera image space, normalized to 0-1) have been widely used as input features in prior work~\cite{he2019device,valliappan2020accelerating,huynh2021imon}. This helps the model calibrate to the relative perspective of the face, which may not be always centered in the camera view. For all of the models discussed in this paper, this base feature set is provided (Figure \ref{fig:architectures}). 

\subsubsection{Eye Crops} 
Perhaps the most conventional input: a cropped RGB patch of both eyes. Using our iris circle estimates, we extract eye patches that are 2.8 times the iris width and 1.4 times the height. Figures \ref{fig:iris_center} and \ref{fig:architectures} provide examples of this cropping. The patches are normalized to a resolution of 500$\times$250 pixels before being fed into the model.

\subsubsection{Screen Thumbnail} 
Our proof-of-concept device, an iPhone 14 Pro Max, has a screen size of 1290$\times$2796 pixels. We apply a Gaussian blur to the screen content, smoothing fine details, which we found to better match the appearance of the screen reflected on the eye. Screen reflections on the eye are very small, often around 10$\times$20 pixels at our iPhone's 4K selfie resolution. Thus, we dramatically reduce processing overhead by downscaling the screen content into a 50$\times$101 pixel thumbnail.

\subsubsection{Reflection Heatmap} 
Using the iris we previously segmented, we crop a reflection search area 0.4 times the iris width and 0.55 times in height, centered on the iris center. We then convolve the screen thumbnail over this cropped region, producing a correlation heatmap. As the size of the reflection varies based on the operating distance and angle between the eyes and the device, we must perform a grid search. We start with a thumbnail width of 0.1 times the iris width, and proportionally scale it up from one to six additional pixels in width and height. We found this range to work well for typical held distances with our 4K images. Each screen scale produces a different correlation heatmap, and we select the one containing the point with the strongest match score. We then downscale the winning heatmap to a standard $40\times55$ pixel input. Note we do not have to perform a grid search for rotation, as the camera and screen (and its reflection) are fixed with respect to one another (i.e., always aligned). 

\subsubsection{Reflection Vector}
Our winning reflection heatmap encodes both the center of the screen reflection, and the best fitting size of the screen reflection convolution filter. Thus, in practice, we actually have a bounding box of the screen reflection. We use this data to produce a vector from the iris center (i.e., center of our heatmap) to the center of the screen reflection. We then normalize this vector by the width and height of the screen reflection bounding box, from -0.5 to 0.5. Thus, if a user is looking at the top-right corner of the screen, the normalized vector would theoretically be around (0.5,~0.5), and when looking at the bottom-left corner, we would roughly expect (-0.5,~-0.5).  

\subsection{Model Architecture}
\label{sec:model_architecture}
Figure~\ref{fig:architectures} illustrates the model architecture we employ. This design closely follows seminal prior work~\cite{valliappan2020accelerating}, but with some key modifications. In all cases, we pass in our \textit{Eye Bounds} data (Section \ref{sec:eyebounds}), which is processed through a set of linear layers with ReLU activations (output dim = 16). Left and right \textit{Eye Crops} are fed into a shared MobileNetv4 backbone (small, e1200, r224, 3.8 M parameters) \cite{qin2024mobilenetv4}, extracting features with an output dimension of 1280 with average pooling. The \textit{Screen Thumbnail} is passed through two lightweight CNN layers, producing a 9600-dimensional flattened embedding. 

The \textit{Reflection Heatmap} is processed through an identically structured CNN, outputting a 4160-dimensional flattened embedding. Both embeddings are then reduced to 256 dimensions through a fully connected layer before concatenation. The \textit{Reflection Vector} is concatenated directly. We note there are cases when the screen reflection cannot be found on the cornea, almost always due to the screen contents being uniformly very dark. When this condition occurs, we simply mask out the \textit{Reflection Vector} input. Finally, all embeddings are concatenated and passed through a multi-layer perceptron to generate the gaze location estimate. 

\subsection{Performance} 

We carefully designed our model to run efficiently on hardware-constrained mobile devices. We benchmarked its inference performance on an Apple iPhone 14 Pro Max (with Apple's A16 Bionic chip). Our best performing model, \textit{Eye Crops + Reflection Vector}~(Figure~\ref{fig:study1_result1}), takes an average inference time of 0.78 ms per gaze prediction (converted to CoreML; standard deviation of 0.03 ms and a memory footprint of 5.1 MB). For preprocessing, we benchmarked our Python pipeline on a MacBook Air (2024, M3 chip). MediaPipe face tracking takes ~\textasciitilde10 ms on average, and iris center correction --- including the GrabCut contour segmentation, RANSAC outlier filtering, and geometric circle fitting --- takes \textasciitilde60 ms. Finally, convolving the \textit{Screen Thumbnail} over the inner-iris, which produces both \textit{Reflection Heatmap} and \textit{Reflection Vector}, takes ~\textasciitilde10 ms. In total, our proof-of-concept system performs end-to-end gaze estimation in roughly 100 ms per frame, corresponding to a throughput of 10 FPS. A commercial-level, compiled implementation (rather than interpreted Python) would likely offer speed improvements.

\subsection{Model Training Protocol}

In our subsequent studies, we employ a leave-one-participant-out cross-validation scheme to train and evaluate our models, always testing on unseen users. Our models are trained using the PyTorch deep learning framework. We initialize the MobileNetv4 backbone with ImageNet pretrained weights. Our models are trained to estimate gaze coordinates in screen space by minimizing Euclidean loss. Models were trained for 20 epochs using the Adam optimizer, batch size of 64 and a learning rate of 0.00003. Data collection specifics are discussed in their respective sections.

\section{Screen Content Dataset}
\begin{figure}[t]
    \centering
    \includegraphics[width=0.84\linewidth]{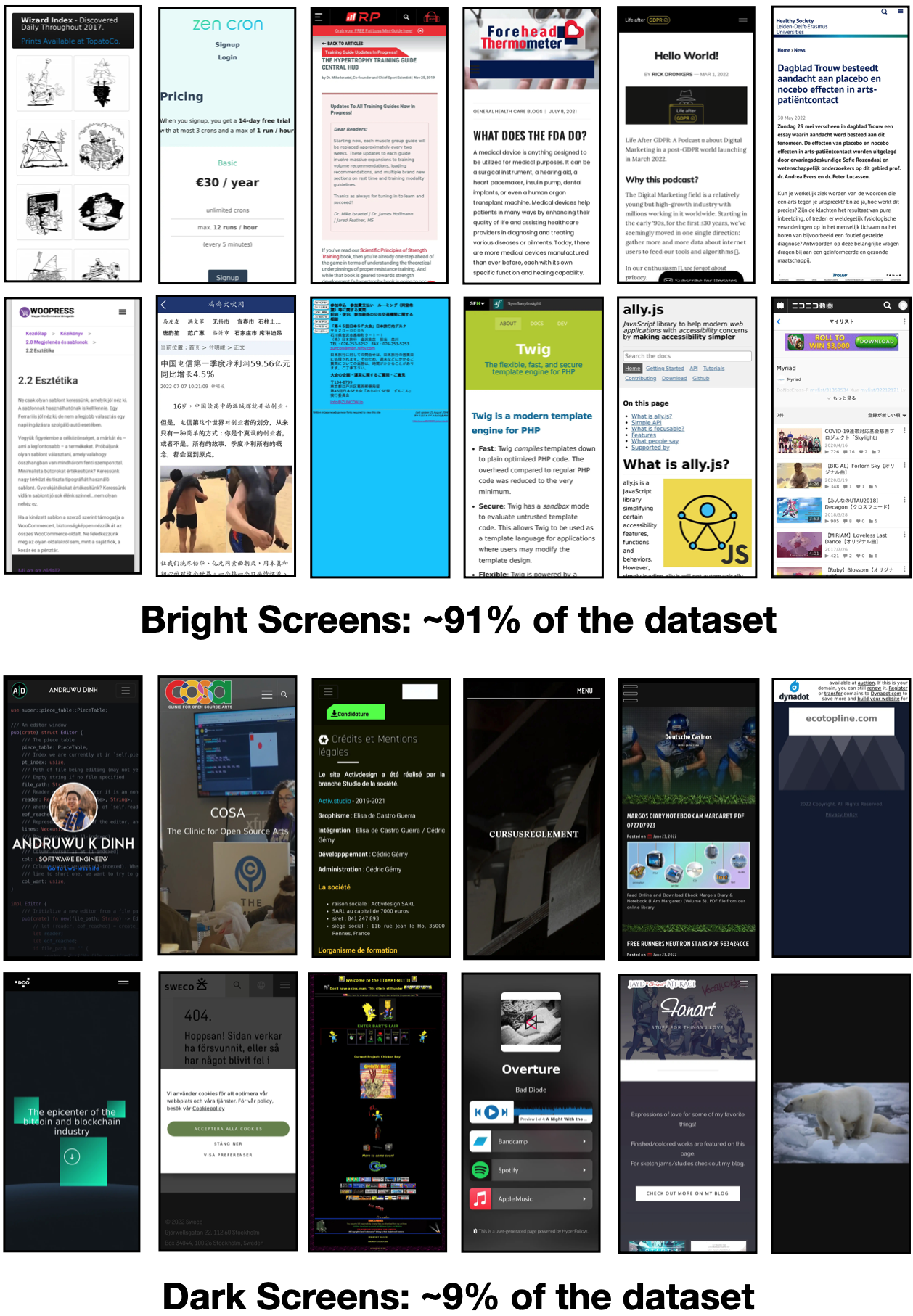}
    \caption{Example screens drawn from the WebUI~\cite{wu2023webui} dataset, which we used in our data collection. Approximately 9\% of the dataset contains predominantly dark screens, which are more challenging for our approach.}.  
    \label{fig:screen_brightness}
    \Description{This figure shows example screen images from the WebUI dataset. Most screens are bright, while a smaller subset uses dark or low-luminance designs. The imbalance highlights the challenge of handling dark screens in reflection-based gaze estimation.}
\end{figure}

As our technique relies on the reflection of screen content in the user's eye, it was paramount to utilize a large and diverse dataset of realistic digital content. For this purpose, we selected WebUI~\cite{wu2023webui}, a public dataset of \textasciitilde300K mobile web pages. Crucially, this contains a representative set of bright and dark interfaces (approximately a ratio of 9:1), as well as dense and sparse arrangements of page-level content. Some example screens can be seen in Figure~\ref{fig:screen_brightness}.

\section{Data Collection Software} 
\label{sec:StudySoftware}

\begin{figure}[t]
	\centering
	\includegraphics[width=\linewidth]{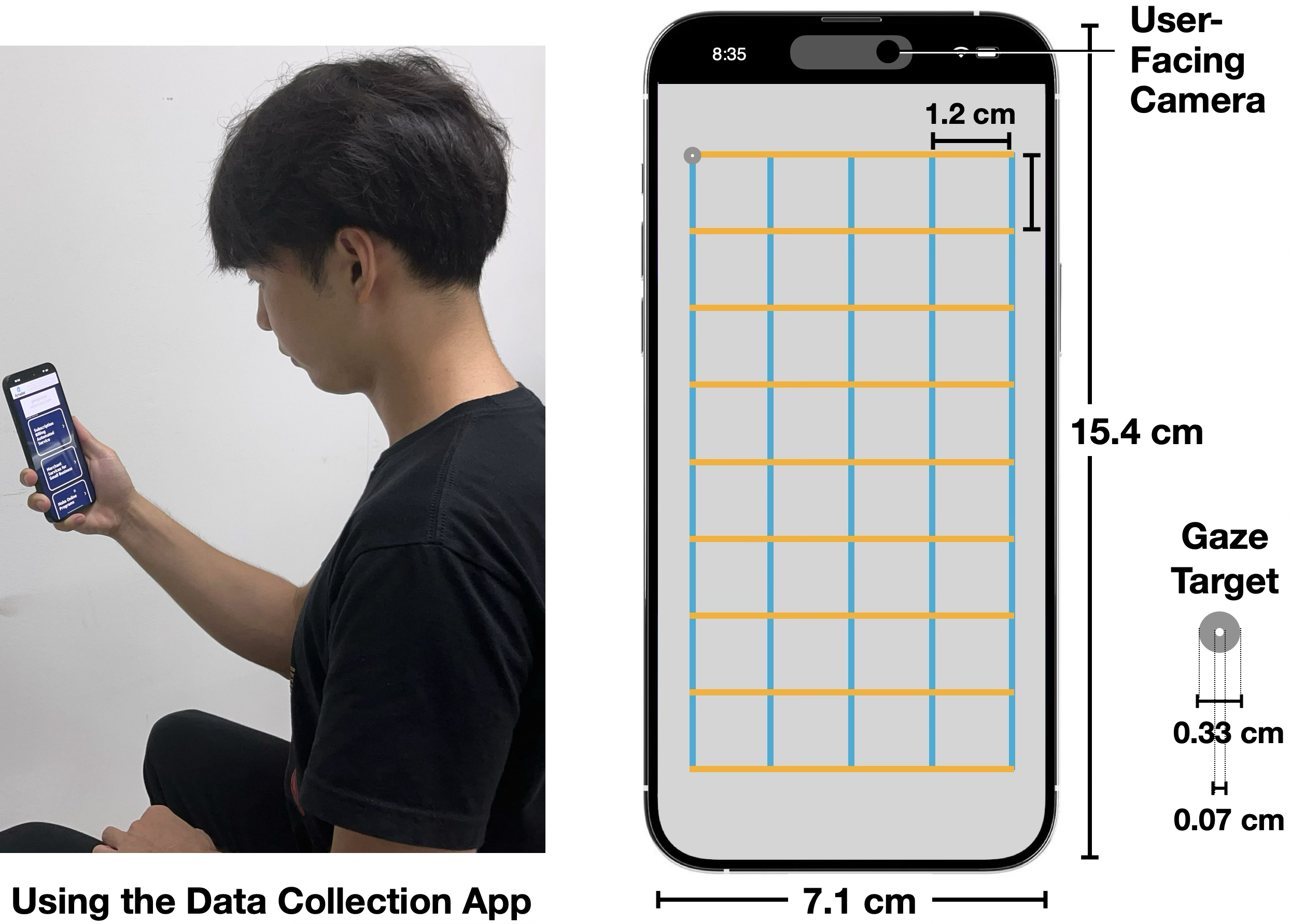}
	\caption{Illustration of the gaze target trajectories used in our data collection software, running on an iPhone 14 Pro Max. Orange and blue lines show the extent and spacing of our horizontal and vertical paths. The on-screen gaze target is shown at the same scale.}
	\label{fig:data_collection_software}
    \Description{This figure illustrates the data collection procedure. A participant looks at a smartphone while a gaze target moves across the screen in predefined horizontal and vertical paths. The trajectories ensure dense coverage of the display area during calibration-free data collection.}
\end{figure}

We created a custom iOS app for data collection, seen in Figure~\ref{fig:data_collection_software}. For ground truth gaze position, users were instructed to follow a small target that animated smoothly across the screen at a constant speed of 100~pixel/s (1.65~cm/s). On our test device, an iPhone 14 Pro Max (see Section \ref{sec:hardware}), we used 9 horizontal paths and 5 vertical paths, equally spaced apart, creating a dense cross-hatch of gaze target paths. We dropped the first 500 ms of data from each path animation to allow participants' smooth pursuit to stabilize. The dropped portions were later covered by counter-moving paths. In total, a horizontal path took 3.5 seconds to animate from one side of the screen to the other, and a longer vertical path took 7.1 seconds. The gaze target's motion was slow and easily tracked by the user, permitting smooth pursuit~\cite{robinson1965mechanics}, ideal for capturing large volumes of data. To further facilitate targeting and give participants breaks, each path trial was separated by a rest screen; participants had to tap a button to continue to the next path. The start of the next gaze target path was previewed on this rest screen so that participants could begin the path trial already fixated on the target. 

We iterated through many gaze target designs to ensure there was no contamination of our captured data (e.g., if using a green dot~\cite{valliappan2020accelerating}, even at low resolution, there will be a small, human-imperceptible increase in green values in the captured pixels of the corneal reflection, which a model could learn). A colored target design was commonly used in previous works \cite{valliappan2020accelerating,DBLP:conf/icmi/ArakawaG0A22,krafka2016eye}, but we believe it to be dangerous for our model. Our final design is a small dark gray circle (0.33 cm) with a central white dot (0.7 mm), similar to the AB target tested in \cite{thaler2013best}. When low-pass filtered (both by insufficient resolution of the optical system and natural blurring/scattering on the surface of the eye) the target is invisible to our model (i.e., below the noise floor). 

To capture data for a wide and realistic variety of possible screen content, our application randomly draws images from the aforementioned  WebUI~\cite{wu2023webui} dataset. A new screen is shown every 400 ms. To further increase content diversity, while also mitigating any possible synchronization issues between the screen content and camera capture, we apply a 200 ms cross-dissolve when switching to new content. 

Our app opens the user-facing camera at 4K (8.3 MP) 60 FPS, but only records frames at 20 FPS. At the same instant, it logs the ground-truth gaze target and also what content was displayed on the screen, including partial cross-dissolves between two screens. We describe trial ordering, repeats, and other study-specific details in the study sections below. 

\section{User Study}

The central question of this work was whether a gaze estimation model would perform better when given knowledge of the screen content (vs. having to rely entirely on RGB appearance). As discussed in Section \ref{sec:EncodingStrategies}, we consider four types of information: 1) \textit{Eye Crop} (i.e., a conventional baseline), 2) \textit{Screen Thumbnail}, 3) \textit{Reflection Heatmap}, and 4) the \textit{Reflection Vector} (as noted previously, \textit{Eye Bounds} data was given to all models as base features). While all of these sources seemingly contain useful information, it was unclear which would perform strongest, nor what combinations of data might be particularly potent. Thus, in this study, we perform a head-to-head evaluation of models operating on different inputs, and use the results to discuss larger trends.

\begin{figure*}[b]
	\centering
	\includegraphics[width=0.90\linewidth]{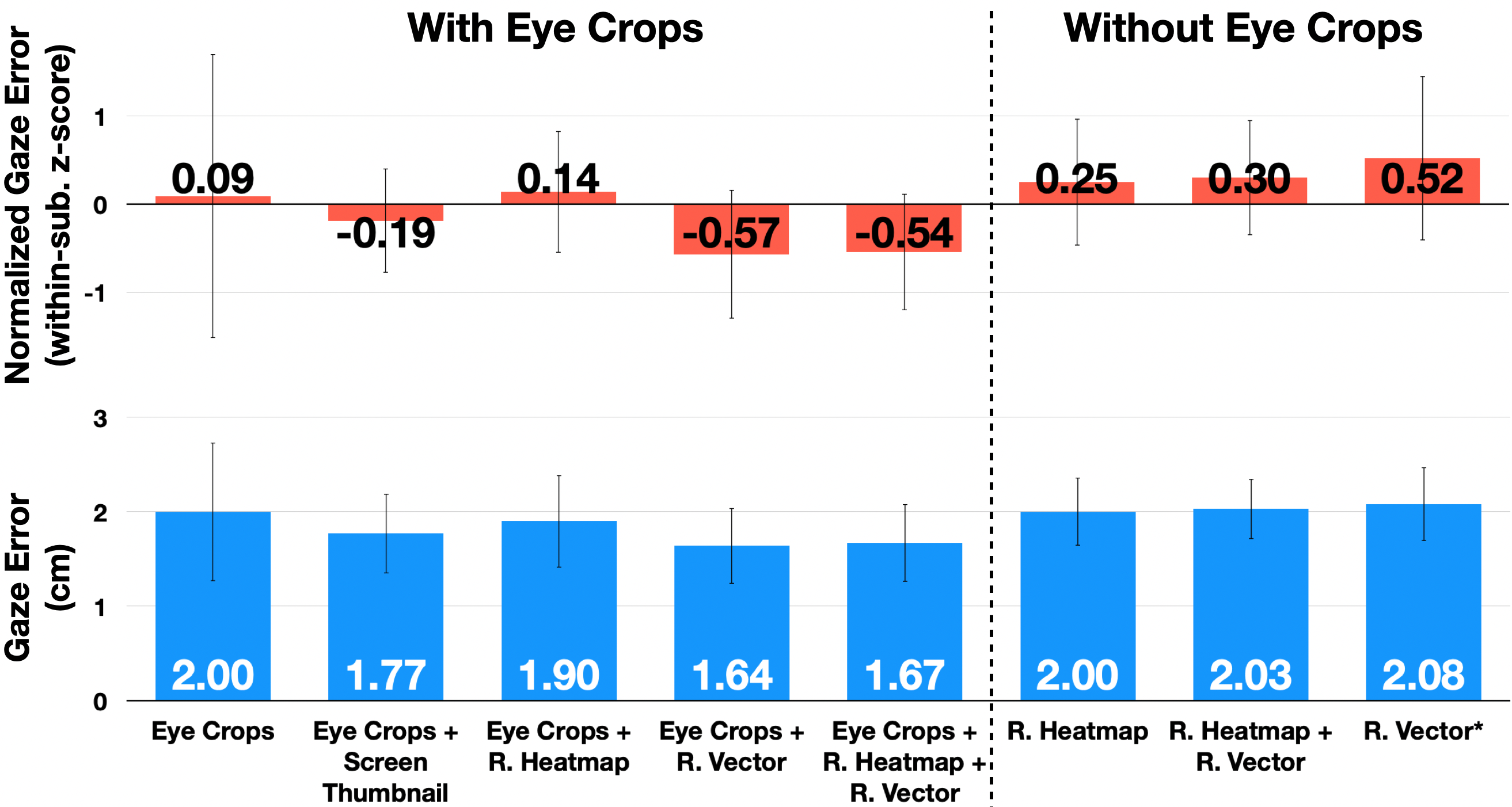}
	\caption{Calibration-free gaze estimation accuracy across eight encoding approaches (note all models are given \textit{Eye Bounds} data as input). Gaze error reported in blue; within-participant normalized gaze error plotted in orange. Lower values are better. \textit{Reflection Vector}* is evaluated only on a subset of gaze instances (153K of 171K; 89\%). Error bars are $\pm 1$ standard deviation.}
	\label{fig:study1_result1}
    \Description{This figure compares gaze estimation accuracy across different input configurations. Results are shown both as absolute error and normalized within-participant performance. Methods that incorporate screen-content-derived features achieve lower error than eye-only baselines.}
\end{figure*}

\subsection{Participants \& Data Collection Procedure}
We recruited 22 participants for this study (12 male, 10 female, age range 19-42, mean age 28), which lasted approximately 30 minutes, including breaks. The study was conducted in a small office with a mean luminance of 213 lux (i.e., typical office lighting). To increase ecological validity, we let participants adjust the iPhone's screen brightness themselves to a "bright but comfortable setting" (mean participant-selected screen brightness was 55\%, SD=11\%; no participant set below 30\% brightness). We also placed no constraints on posture or head position beyond a maximum face-to-phone distance of 45 cm. Participants were not forced to artificially hold their head still; they completed the study with natural head and hand poses, as well as movement.

An iPhone 14 Pro Max ran our study software (Sections \ref{sec:hardware} and \ref{sec:StudySoftware}). One session of data collection consisted of each animated gaze path and direction appearing once (9 horizontal and 5 vertical paths), with the user following along with their gaze of smooth pursuit. As noted previously, data was continuously captured during the path animations. One session of data collection yielded \textasciitilde1.3K data points. Participants completed 6 sessions in total: 3 while sitting and 3 while standing, in alternating order. Between sessions, they were asked to place the phone down and lift it back up again, introducing variation in phone-head distance and angle, hand grip, body posture, and head pose. After completing all 6 sessions, each participant yielded \textasciitilde8K data points. In total, our 22 participants produced \textasciitilde177K eye tracking instances, of which 5.6K blink instances (\textasciitilde3\%) were excluded from subsequent analyses. 

We note that we did not capture data for participants while wearing glasses --- this is an important special condition that requires additional considerations (and other seminal papers have similarly set this potential confound aside \cite{valliappan2020accelerating,huang2017screenglint}). Glasses have many styles, optical power, reflective coatings, etc., and it was beyond the scope of this initial work to collect a dataset that could represent this population. Instead, we leave this factor for future work, focusing on our core hypotheses. We did, however, recruit participants who wore glasses, but all of them were able to focus on a moving stimulus without the aid of glasses. 

\subsection{Baseline Condition}

To assess the relative performance of our technique, we include a baseline method. For this, we use our \textit{Eye Crops} model, which is widely used as a competitive baseline in recent gaze estimation work~\cite{valliappan2020accelerating,kong2021eyemu,huynh2021imon}. 

We also considered two other baselines: a full-face appearance model~\cite{zhang2017s} and a head-pose augmented model~\cite{valenti2011combining}. Through ablation tests, we checked whether adding a full-face crop and/or full 468 MediaPipe head-pose landmarks (i.e., complete head-pose information) to our \textit{Eye Crops} model could improve accuracy. We found these two approaches performed the same or worse than the regular \textit{Eye Crops} model, and so we selected the latter as our "gold standard" baseline (plotted against our HiFiGaze models' performance in Figures \ref{fig:study1_result1}-\ref{fig:study1_analysis2}).

\subsection{Results \& Discussion}
\label{sec:mainResults}

Figure~\ref{fig:study1_result1} presents our study's main results. All models were trained with leave-one-participant-out cross-validation and  evaluated on unseen users.

On the bottom plot, we report mean gaze tracking error across our 22 participants. Note the substantial error bars (standard deviation) due to performance variation across users --- the best participant had a mean error of 1.54 cm when averaged across the eight models, while the worst-performing participant had a mean error of 2.51 cm. To control for this cross-user performance variation, we also provide within-participant normalized performance and plot the z-scored means for each input scheme in the top plot.

Starting first with the baseline \textit{Eye Crops} model --- a classic appearance-based method --- we find a mean error of 2.00 cm. This accuracy is in line with similarly architected prior work~\cite{huang2017screenglint,valliappan2020accelerating}, before any per-user calibration is applied, and lends credence that our model architecture (which is similar, but adapted from prior work for our purposes) is competitive and performing well. 

Next, we see that all models utilizing eye crops data \textit{plus} screen content knowledge improve in accuracy (conditions \textit{Eye Crops + Screen Thumbnail}, \textit{Eye Crops + Reflection Heatmap}, \textit{Eye Crops + Reflection Vector}, and \textit{Eye Crops + Reflection Heatmap + Reflection Vector}) over the \textit{Eye Crops} baseline. This is strong evidence that screen content knowledge is useful to the model. Using a linear mixed effects model, we examined the contribution of each information source while accounting for inter-participant variability (fixed effects: \textit{Eye Crops}, \textit{Screen Thumbnail}, \textit{Reflection Heatmap}, and \textit{Reflection Vector}). Likelihood ratio tests revealed that \textit{Eye Crops} information significantly improved model performance (\textit{$\chi^2$(1)} = 18.27; \textit{p} < .001), as did the \textit{Reflection Vector} (\textit{$\chi^2$(1)} = 7.47; \textit{p} < .01). 

The best performing method was \textit{Eye Crops + Reflection Vector}, which performed comparably to \textit{Eye Crops + Reflection Heatmap + Reflection Vector}. Both models significantly outperformed \textit{Eye Crops} (both \textit{p} < .005, linear mixed-effect model). \textit{Eye Crops + Screen Thumbnail} achieved the next best performance and also significantly outperformed \textit{Eye Crops} (\textit{p} < .05). It is likely optimizing itself using features extracted from the \textit{Screen Thumbnail}, such as the overall lightness/darkness, the presence of icons and other identifying markers, etc. The \textit{Eye Crops + Screen Thumbnail} model is particularly attractive because it requires virtually no processing (i.e., no extra processing to segment the screen's reflection on the eye), and can work on any input. 

\begin{figure}[t]
    \centering
    \includegraphics[width=\linewidth]{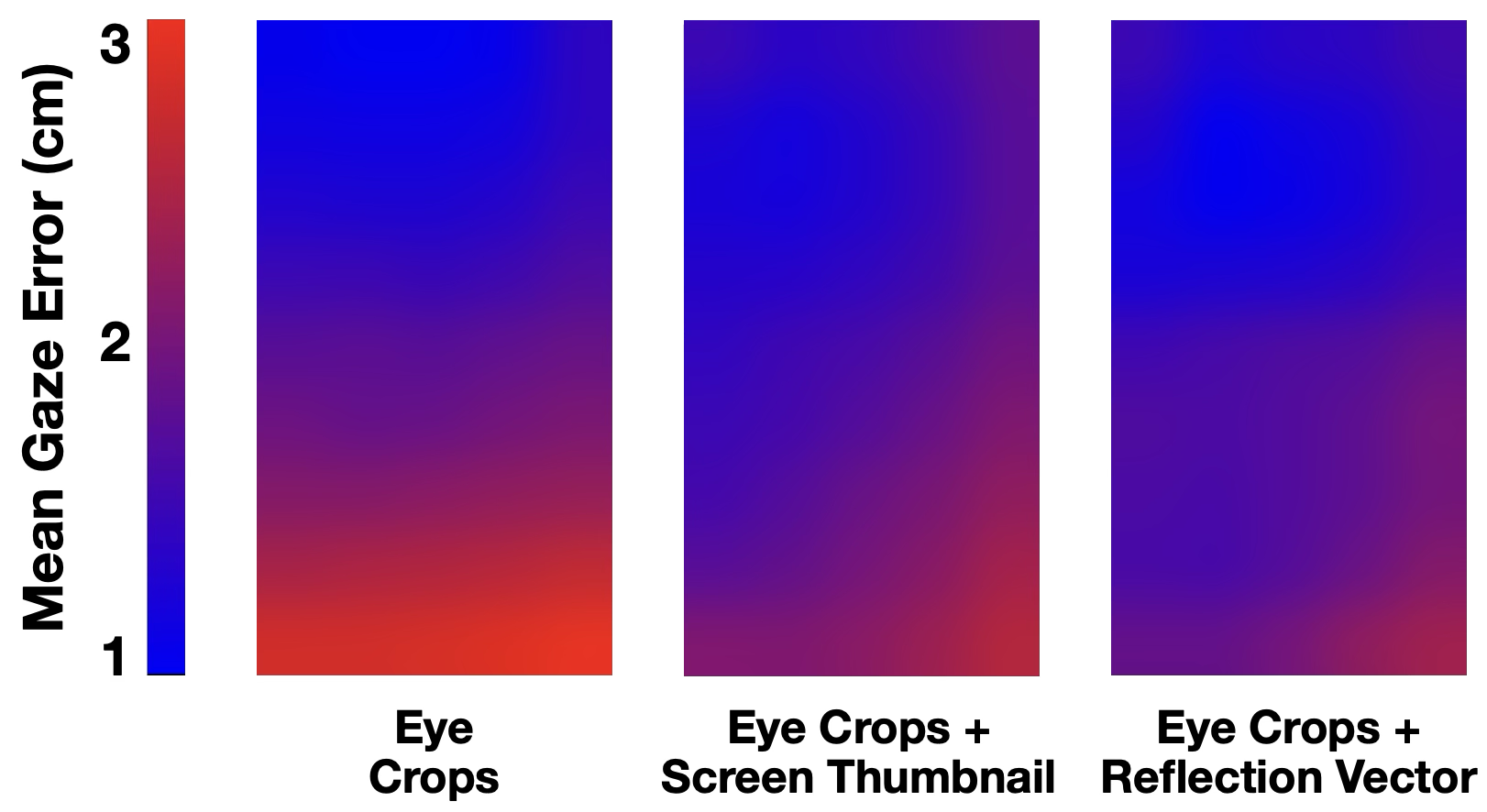}
    \caption{Spatial heatmaps of gaze error across the screen.}
    \label{fig:study1_analysis1}
    \Description{This figure visualizes spatial gaze error distributions over the phone screen. Heatmaps show higher error near the lower screen region for baseline methods. Incorporating screen knowledge reduces error, particularly in these challenging areas.}
\end{figure}

We can also see in our results that \textit{Reflection Heatmap} data alone (without the \textit{Eye Crops}) offers similar performance as \textit{Eye Crops} (2.00 vs. 2.00 cm mean error, and with less variance), a surprising and informative result. The 2D heatmap is a rich data source containing other visual cues that the more fully processed \textit{Reflection Vector} cannot. Of note, our grayscale correlation heatmaps obfuscate user biometric data (most notably the complex iris pattern, used in some authentication systems), and could potentially offer a more privacy-preserving route to enabling cloud-based gaze tracking models with less privacy risks.

On the right in Figure~\ref{fig:study1_result1}, we include results for a \textit{Reflection Vector} only model, which is just four floating point numbers (plus the base \textit{Eye Bounds} features). We mark this result with an asterisk because it is our only model that operates on a subset of our dataset. When the screen reflection is not found on both eyes --- e.g., very dark screen content leaving no segmentation clue on the black iris --- there are no values to pass to the model, and so we skip the gaze instance. As such we were only able to run this model on 89\% of our corpus (with the missing 11\% likely to be the most challenging reflections), and thus it is not able to offer an apples-to-apples comparison with our other methods. Nonetheless, we report this accuracy, as it shows, at least for a majority of screen content, that even this very simple encoding is reasonably accurate (when it is able to run).

\begin{figure}[t]
    \centering
    \includegraphics[width=\linewidth]{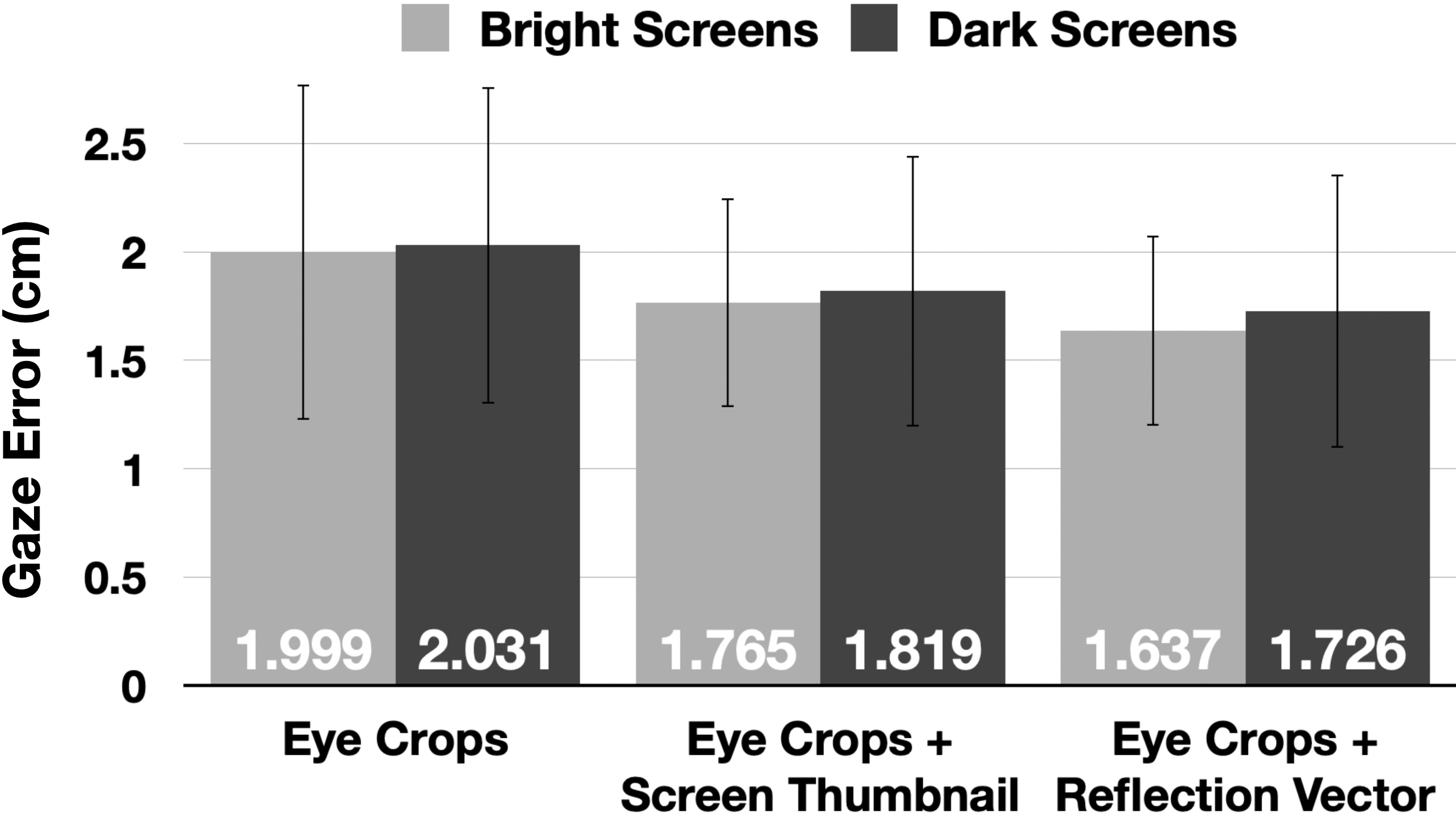}         
    \caption{Mean gaze error broken out by method and instances with bright and dark screen content. Note that the \textit{Eye Crops} condition is equivalent to prior appearance-based methods. See bright and dark screen examples in Figure \ref{fig:screen_brightness}.}
    \label{fig:study1_analysis2}
    \Description{This figure compares gaze estimation performance under bright and dark screen conditions. Errors are generally higher for dark screens. However, models using screen knowledge consistently outperform the baseline in both cases.}
\end{figure}


For further analysis, we created heatmaps of error across the screen surface in Figure \ref{fig:study1_analysis1} for three models: \textit{Eye Crops}~(baseline), \textit{Eye Crops + Screen Thumbnail}~(our most flexible model), and \textit{Eye Crops + Reflection Vector}~(our most accurate model). As is readily apparent, there is a trend of increased error as targets get lower on the screen (a result consistent with prior work~\cite{valliappan2020accelerating,krafka2016eye}). Notably, these lower-screen offsets are alleviated with the addition of screen knowledge features, suggesting that these signals help maintain accuracy across a wider range of the screen, particularly at locations farther from the camera at the top of the device.

We also hypothesized that screen content brightness would have an effect on gaze estimation performance. Our analysis revealed that darker screens led to slightly higher estimation errors across all models~(Figure~\ref{fig:study1_analysis2}). The \textit{Eye Crops + Screen Thumbnail} model had a \textasciitilde3\% increase in error (1.77 to 1.82 cm) and the \textit{Eye Crops + Reflection Vector} model had a \textasciitilde5\% increase (1.64 to 1.73 cm) in error. Notably, our models consistently outperformed the baseline even when the screen content was dark, when specular reflections are minimal. In fact, our best model (\textit{Eye Crops + Reflection Vector}) achieved lower error on dark content (1.73~cm) than the baseline model did on bright content (2.00~cm). We believe this robustness stems from our pipeline's ability to lock onto small bright details, like logos and text, while the general context of screen content (i.e., knowing the screen is mostly dark) also aids the model.

\section{Supplemental Study: Camera Location}
A spatial analysis of error in our user study revealed an increase in gaze estimation error towards the bottom of the screen (Figure~\ref{fig:study1_analysis1}). Upon closer inspection of data, it was apparent that bottom targets were sometimes occluded by the upper eyelid and eyelashes, impacting our screen-content-derived features and thus model accuracy. 

One possible solution to this innate issue (faced by most, if not all vision-based gaze methods) is to relocate the camera downwards, to the bottom of the phone. Due to the vertical asymmetry of human eyes, a lower view point can reduce both eyelid and eyelash occlusion. To quantitatively assess this accuracy benefit, we ran a supplemental study. While this camera placement is atypical, it is certainly not impossible to implement, and perhaps could utilize behind-screen camera technology~\cite{cha2022} so as not to interfere with the continuity of the display. 

\begin{figure}[b]
	\centering
	\includegraphics[width=1\columnwidth]{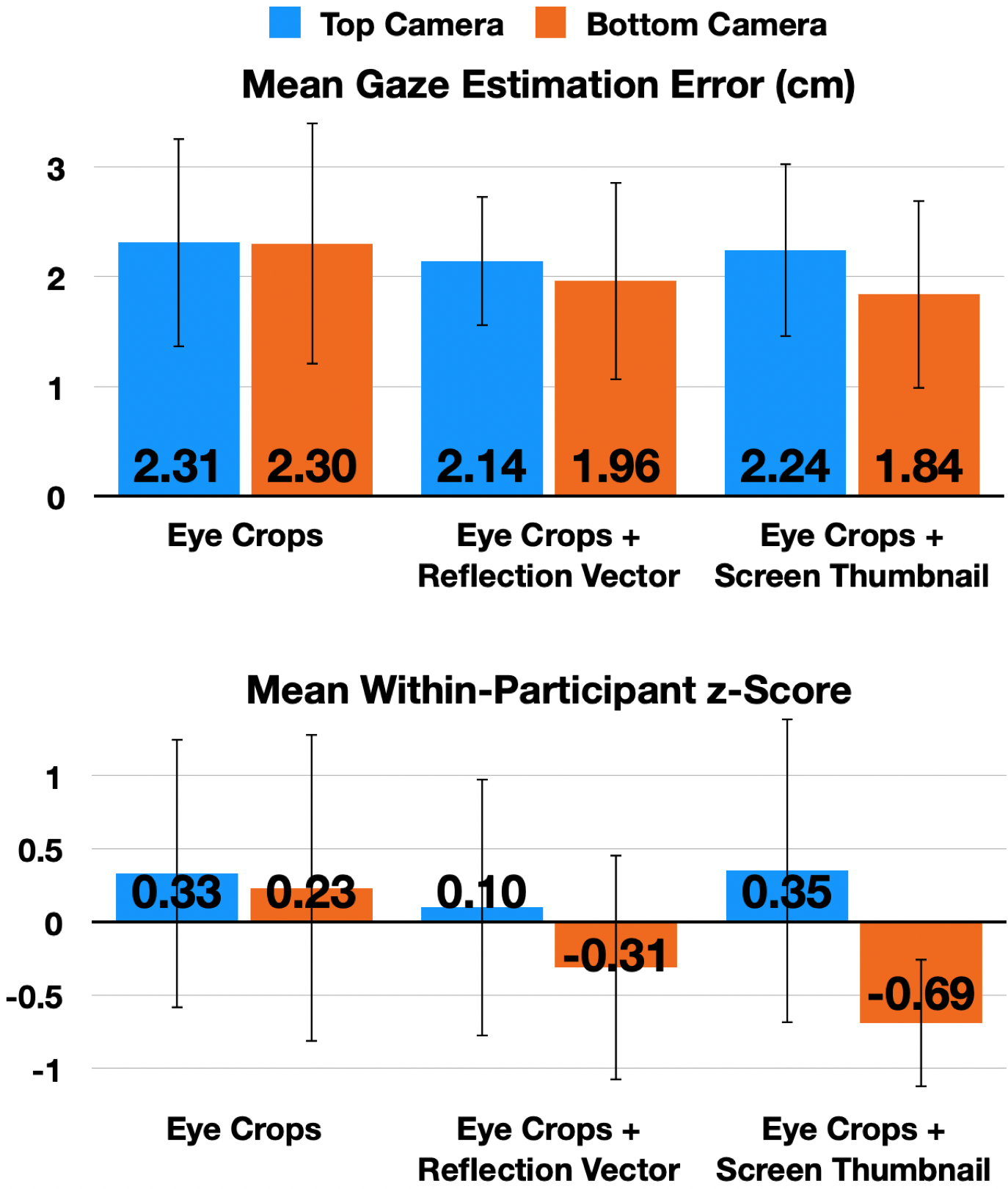}
	\caption{Calibration-free gaze estimation error for top and bottom camera location. Error bars are $\pm 1$ standard deviation.}
	\label{fig:study2_result1}
    \Description{This figure compares top-camera and bottom-camera smartphone configurations. Bottom-camera placement results in lower gaze estimation error across methods. The improvement highlights the impact of camera geometry on appearance-based gaze tracking.}
\end{figure}

\subsection{Study Apparatus, Procedure, \& Participants}
The study apparatus, software, and procedure were identical to our main study, except that we flipped the iPhone upside-down, such that its user-facing camera was now at the bottom of the phone. The screen display was also inverted so that the original bottom aligned with the top of the screen. Using the pool of participants from our main study, we re-recruited 10 participants (5 male, 5 female; age range 23-33, mean age 28). This meant that we already had matched data for these ten participants with an upper camera location, allowing us to run a within-participants comparison.

\subsection{Results \& Discussion}

Figure~\ref{fig:study2_result1} presents the results of this supplemental study, evaluated using a leave-one-participant-out scheme, always testing on unseen users. Note that the participant pool was small (n = 10), such that estimation errors were generally higher than the main study due to the reduced amount of training data --- the mean error in the \textit{Eye Crops}, top-camera condition increased to 2.31 cm, compared to the 2.00 cm in the earlier full study with 22 participants. Since the purpose of this study is to observe the trend \textit{within} participants, we interpret the results in terms of relative performance rather than absolute numbers.

The top plot of Figure~\ref{fig:study2_result1} shows that the advantage of bottom-camera placement is evident in both of our models that leverage screen reflections: \textit{Eye Crops + Reflection Vector} improves by \textasciitilde8\% (from 2.14 to 1.96 cm), and \textit{Eye Crops + Screen Thumbnail} improves by \textasciitilde18\% (from 2.24 to 1.84 cm). Meanwhile, the baseline \textit{Eye Crops} model shows no measurable improvement in the bottom-camera setting. This suggests that our proposed models, which leverage screen content knowledge, suffer more from occlusions of the eye caused by the upper eyelid and eyelashes than the conventional \textit{Eye Crops} model, leaving room for accuracy gains when the camera is placed at the bottom in practice.

The z-scored means seen in Figure~\ref{fig:study2_result1} (eliminating between-participant variations) reveal this pattern more distinctly. Both \textit{Eye Crops + Reflection Vector} and \textit{Eye Crops + Screen Thumbnail} show notable error reductions when the camera is moved from top to bottom, whereas the conventional \textit{Eye Crops} does not. Taken together, the consistent trend confirms that bottom-camera placement offers an advantage, particularly for our proposed approach.

\section{Summary of Findings}
\label{sec:summary}
Our studies revealed four main findings, summarized below: \\


\noindent\textbf{1. Using screen content knowledge systematically improves gaze estimation performance over conventional appearance-based methods~\cite{krafka2016eye,valliappan2020accelerating}.}
Our best-performing model (\textit{Eye Crops + Reflection Vector}) reduces gaze error by 18\% (2.00 to 1.64 cm) as compared to the \textit{Eye Crops} baseline. Notably, the \textit{Eye Crops + Screen Thumbnail} model also performs well (12\% error reduction, 2.00 to 1.77 cm). This is particularly attractive as it provides the simplest pipeline with virtually no preprocessing (no iris center estimation, no computation of the reflection vector) and can work on any input. \\

\noindent\textbf{2. The \textit{Reflection Heatmap} data alone (without the \textit{Eye Crops}) offers similar performance to the baseline \textit{Eye Crops} method (2.00 vs. 2.00 cm, with less variance).}
Note that this grayscale correlation heatmap could potentially mitigate privacy risks that occur while transmitting detailed eye imagery between servers. This may provide an opportunity for more secure cloud-based gaze tracking systems, a topic we hope to explore in future work. \\

\noindent\textbf{3. Models that use screen content knowledge outperform the baseline irrespective of screen content brightness.} Dark screen content does not produce salient eye reflections, leading to reduced performance in all models~(Figure~\ref{fig:screen_brightness}). However, our model (\textit{Eye Crops + Reflection Vector}) achieved lower error on dark content (1.73 cm) than the baseline model did on bright content (2.00 cm). We hypothesize that the model is still able to leverage small bright elements for localization (e.g., logos, text, buttons and other widgets), and by passing in a thumbnail of the screen content, the model is able to adapt its behavior knowing if the screen is bright/dark. \\


\noindent\textbf{4. One source of error was the partial occlusion of corneal screen reflection by the users' upper eyelids and eyelashes, especially during downward gaze}. Our supplemental study revealed that relocating the camera to the bottom of the device mitigated this issue and yielded accuracy gains.

\section{Limitations \& Future Work}
While our study demonstrates notable performance gains from using screen content knowledge on consumer devices, important future work remains.

\begin{figure*}[t]
	\centering
	\includegraphics[width=1.0\linewidth]{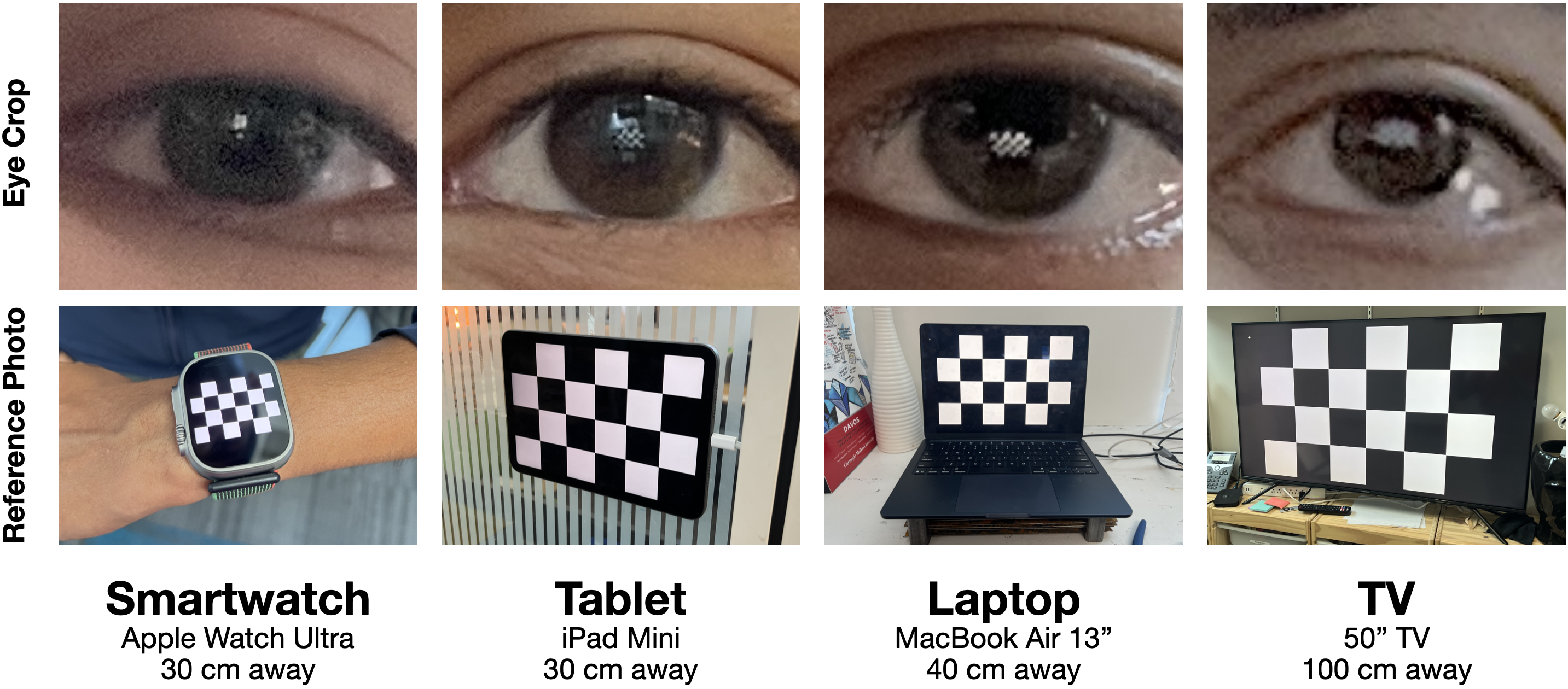}
        \caption{Various everyday computing devices where the HiFiGaze eye tracking method is applicable.}
	\label{fig:ExampleDevices}
    \Description{A four-column comparison showing eye reflections from different display types. The top row presents close-up eye crops with checkerboard patterns reflected in the pupil. The bottom row shows the corresponding reference photos of the displays: an Apple Watch Ultra smartwatch at 30 cm, an iPad Mini tablet at 30 cm, a 13-inch MacBook Air laptop at 40 cm, and a 50-inch TV at 100 cm. The checkerboard reflection grows smaller and less distinct as viewing distance and display size increase.}
\end{figure*}

First, broader data collection under more diverse conditions will be necessary to fully capitalize on our approach and achieve reliable real-world performance. In our present study, participants were free to adjust their hand and head positions between recordings, and also put down and pick up the phone between sitting and standing sessions, which added natural variations in distance, posture and grip. However, this falls short of covering the full diversity of phone-user poses in the wild. 
Future work will need to examine more dynamic conditions, including walking~\cite{tomasi2016mobile, DBLP:conf/icmi/ArakawaG0A22}, uncontrolled hand and head motions~\cite{krafka2016eye}, vehicular sway~\cite{trosterer2014eye}, and postures such as lying down~\cite{DBLP:conf/icmi/ArakawaG0A22, huang2017tabletgaze}. In addition, our studies were conducted under typical indoor office lighting, but a commercial model will need to incorporate data across a wider range of lighting conditions --- including low light, darkness, and strong side or back lighting --- that may interfere with eye reflections. While it would be rare for any random environmental glint to match the screen thumbnail (as previously discussed), it is more likely that other light would partially occlude the screen reflection. It is also possible that very bright environments could cause the exposure time to be too short for the reflection to register.

Another limitation of our current study is the range of screen-to-eye distances we tested. Our evaluation focused on smartphone usage within typical handheld distances (roughly 30-45 cm). However, users can interact with laptops, desktop monitors, or televisions at viewing distances on the order of meters. Fortunately, these larger devices also have larger eye reflections (some examples shown in Figure \ref{fig:ExampleDevices}), and in future work we hope to extend our method across a range of device types.

As already discussed in Section \ref{sec:mainResults}, our model's performance modestly degrades with darker screen content, as eye reflections become a less useful signal. As noted in our user study, roughly 9\% of our dataset (randomly drawn from the WebUI dataset~\cite{wu2023webui}) was considered to be dark, and thus more challenging. However, interestingly, our best model still outperforms the baseline on both bright and dark instances, suggesting our pipeline could utilize small bright landmarks (e.g., logos, buttons, text headers, and other widgets). It may even be that our model is reverting to essentially an appearance-based approach, but primed with the knowledge that the screen content is dark (since the screen content is passed in as an input to the model). Put simply, the benefits of utilizing eye reflection may diminish with darker content, but there is still value in the model knowing the screen content.

From a methodological standpoint, the accuracy of our approach is tightly coupled to robust iris center estimation. Although our custom correction improved over raw MediaPipe landmarks, we still observed some noise. Since any offset in the iris center directly translates to offset in the reflection vector, improving the stability of iris center detection is a key opportunity. We anticipate that lightweight deep learning iris center detection models, trained on a large, labeled dataset could provide both improved robustness and lower computational cost compared to our current pipeline.

Our participant pool also highlights ecological validity gaps. Of the 22 participants, 21 had brown irises, and only one had blue. While this blue-eyed participant’s performance was close to the group average, our participant pool is not diverse enough to say with total confidence we can generalize across iris colors. Similarly (and inline with other prior work~\cite{valliappan2020accelerating,huang2017screenglint}), we excluded users wearing glasses, which constitute a large portion of everyday device users. Glasses introduce secondary reflections, optical distortions, and potential attenuation of the corneal reflection, all of which would challenge our method. There is no reason our method cannot overcome this, and it may be that screen content knowledge can be of great help, but a future investigation should be undertaken. We also did not explicitly quantify the effect of individual eyelash morphology, which appears to play a meaningful role in occlusion of the reflection for downward gaze. 

Beyond demographic factors, we observed substantial variability in reflection quality across individuals, even under matched lighting and device conditions. This may relate to differences in tear film stability (the transparent liquid layer that coats the surface of the eye) or corneal shape. For example, participants with astigmatism (8 of 22 in our study) had irregular corneal curvature, which likely blurred or distorted their corneal reflections relative to others. These physiological sources of variance highlight the difficulty of building a one-size-fits-all model, and suggest that either adaptive modeling or lightweight per-user calibration may be necessary to reach consistently high accuracy.

\section{Conclusion}
We have introduced a new approach to estimate gaze on everyday computing devices that leverages a previously underutilized signal: the reflection of a device’s own display in the user’s eyes. By combining high-resolution, user-facing camera imagery with knowledge of the screen content, our method enables robust segmentation of the screen reflection and uses its position and extent as a reliable proxy for screen-relative gaze. Through our evaluations, we demonstrated that this approach improves accuracy up to \textasciitilde18\% over traditional appearance-based techniques, achieving calibration-free, icon-sized gaze precision on a contemporary smartphone. As devices continue to ship with ever higher-resolution user-facing cameras, reflection-based methods could provide a path toward practical, widely deployable gaze tracking. We believe this opens up new opportunities for gaze-aware interfaces, accessibility applications, and everyday attention-sensing interactions on the devices people already own.

\begin{acks}
This work was supported in part by the IITP (Institute of Information \& Communications Technology Planning \& Evaluation) - ITRC (Information Technology Research Center) grant funded by the Korean government (Ministry of Science and ICT) (IITP-2026-RS-2024-00436398). 
\end{acks}

\balance
\bibliographystyle{ACM-Reference-Format}
\bibliography{main}

\newpage    
\appendix
\end{document}